\documentclass[twocolumn,showpacs,preprintnumbers,amsmath,amssymb,superscriptaddress]{revtex4}

\usepackage{graphicx}
\usepackage{dcolumn}
\usepackage{bm}
\usepackage{CJK}

\begin{document}

\title{Chaoticons described by nonlocal nonlinear Schr\"{o}dinger equation
}

\author{Lanhua Zhong}
\affiliation{Guangdong Provincial Key Laboratory of Nanophotonic
Functional Materials and Devices, South China Normal
University, Guangzhou 510631, P. R. China}
\affiliation{Physical Science and Technology School, Lingnan Normal University, Zhanjiang 524048, P. R. China}
\author{Yuqi Li}
\affiliation{Shanghai Key Laboratory of Trustworthy Computing, East China Normal University,
Shanghai 200062, P. R. China}
\author{Yong Chen}
\affiliation{Shanghai Key Laboratory of Trustworthy Computing, East China Normal University,
Shanghai 200062, P. R. China}
\author{Weiyi Hong }
\affiliation{Guangdong Provincial Key Laboratory of Nanophotonic
Functional Materials and Devices, South China Normal
University, Guangzhou 510631, P. R. China}
\author{Wei Hu}
\affiliation{Guangdong Provincial Key Laboratory of Nanophotonic
Functional Materials and Devices, South China Normal
University, Guangzhou 510631, P. R. China}
\author{Qi
Guo} \email[Corresponding author's email address:
]{guoq@scnu.edu.cn}
\affiliation{Guangdong Provincial Key Laboratory of Nanophotonic
Functional Materials and Devices, South China Normal
University, Guangzhou 510631, P. R. China}

\date{\today}

\begin{abstract}
 It is shown that the unstable evolutions of the Hermite-Gauss-type stationary solutions for the nonlocal nonlinear Schr\"{o}dinger equation with the exponential-decay response function can evolve into chaotic states. This new kind of entities are referred to as chaoticons because they exhibit not only chaotic properties (with positive Lyapunov exponents and spatial decoherence) but also soliton-like properties (with invariant statistic width and interaction of quasi-elastic collisions).
\end{abstract}

.

\maketitle

\emph{Introduction}--Solitons are self-reinforcing stable localized wave entities that maintain their shapes when they evolve in nonlinear systems, and are caused by a balance between nonlinearity and dispersion in the systems. They have been demonstrated in a large variety of physical systems, including optics, fluid mechanics, particle physics and even astrophysics~\cite{book-Dauxois}. Over the past three decades,
optical solitons~\cite{Kivshar-book-03, Agrawal-book-07-1, Stegeman-science-99, Stegeman-ieee-00, chen-rpp-12, book-Trillo, book-Assanto, guo-book-2015} have been at the forefront of soliton research,
which are modelled by the nonlinear Schr\"{o}dinger equation (NLSE) $i{\partial q}/{\partial t}+({1}/{2}){\partial^{2}q}/{\partial x^{2}}+|q|^{2}q=0$ (for the local nonlinearity)~\cite{Kivshar-book-03, Agrawal-book-07-1, book-Trillo} and its generalized form, the nonlocal nonlinear Schr\"{o}dinger equation (NNLSE) (for the nonlocal nonlinearity)~\cite{Snyder-science-97, Krolikowski-pre-01, Bang-pre-02, Conti-prl-03, Conti-prl-04, Ouyang-pre-06, Deng-josab-07, Buccoliero-prl-07, Xu-ol-05,Dong-pra-10, Rasmussen-pre-05, Ouyang-pra-07, Rotschild-np-06, Hu-pra-08, Skupin-pre-06, Kaminer-ol-07, Rotschild-np-08, Picozzi-prl-11, book-Assanto, guo-book-2015}.
%
The (1+1)-dimensional form of the NNLSE is~\cite{guo-book-2015, Krolikowski-pre-01, Snyder-science-97}
\begin{equation}\label{nnlse}
i\frac{\partial q(x,t)}{\partial t}+\frac{1}{2}\frac{\partial^{2}q(x,t)}{\partial x^{2}}+q(x,t)\int^{\infty}_{-\infty}R(x-\xi)|q(\xi,t)|^{2}d\xi=0,
\end{equation}
where the real positive function $R(x)$ is the (nonlinear) response function, which must be symmetry for the existence of the soliton-like solutions~\cite{Hong-pra-15}.

The nonlocal nonlinearity (the convolution integral) in Eq.~(\ref{nnlse}) means that the wave-induced ``potential'' at a certain spatial point $x$, $V(x,t)=-\int^{\infty}_{-\infty}R(x-\xi)|q(\xi,t)|^{2}d\xi$, is determined not only by the wave $q(x,t)$ at that point but also by the wave in its vicinity.
The stronger the nonlocality, the more extended the wave distribution contributing to the ``potential'' $V$~\cite{Krolikowski-pre-01, Bang-pre-02, guo-book-2013}. Different from the (local) NLSE [when $R(x)=\delta(x)$ in Eq.~(\ref{nnlse})], nonlocality has profound effects on the dynamics of solitons.
For example, the interaction of two nonlocal solitons can have both a long-range mode~\cite{Rasmussen-pre-05, Rotschild-np-06, guo-book-2013} and a short-range mode~\cite{Ouyang-pra-07, Hu-pra-08, guo-book-2013}, but two local solitons interact with each other only in a short-range one~\cite{Stegeman-science-99, guo-book-2013}; and the NNLSE [Eq.~(\ref{nnlse})] can support the multi-hump solitons with the Hermite-Gauss-type (HGT) profiles~\cite{Deng-josab-07, Buccoliero-prl-07, Xu-ol-05, Dong-pra-10}, but the NLSE admits only the single-hump solitons~\cite{Agrawal-book-07-1}.
However, the NNLSE may not guarantee the existence of all high order HGT-solitons. The response function also plays an important role. The NNLSE with the Gaussian response function 
 can support the HGT-solitons without upper threshold of the hump-number~\cite{Xu-ol-05, Deng-josab-07, Buccoliero-prl-07}. Contrastively, the NNLSE with the exponential-decay response function (EDRF)~\cite{Krolikowski-pre-01} only admits of the HGT-solitons with the hump-number less than 5~\cite{Xu-ol-05}. The crucial difference between such two kinds of response functions is~\cite{guo-book-2013, guo-book-2015} that the former is non-singular and the potential $V$ can be simplified to a quadratic form in the limit of strong nonlocality, while the latter that can describe physically real materials is singular and the corresponding NNLSE cannot be generally reduced to a linear Snyder-Mitchell mode~\cite{Snyder-science-97}.

On the other hand, the NNLSE given by Eq.~(\ref{nnlse}) is non-integrable~\cite{chen-rpp-12, Picozzi-prl-11}. In a non-integrable nonlinear system, chaos often appears. Chaos is generally agreed to denote the aperiodic long-term behavior of a bounded deterministic system that exhibits sensitive dependence on initial conditions.
 And the most common criterion for chaos is a positive Lyapunov exponent, which means that two initially arbitrarily close trajectories in phase space diverge exponentially in time~\cite{book-Sprott, book-Schuster, Lissauer-rmp-99}.

In this letter, we investigate the evolution of the (1+1)-dimensional NNLSE with the EDRF for the initial inputs of the HGT stationary solutions~(SSs). As has been mentioned~\cite{Xu-ol-05}, the HGT-SSs with the hump-number more than 4 always evolve unstably. We, however, find that such an unstable evolution of every HGT-SS can develop into a chaotic state, which is characterized by the positive Lyapunov exponent and spatial decoherence. Moreover, it also exhibits the soliton-like properties: the invariant statistic width during the evolution and the quasi-elastic collisions during the interaction. Therefore, we refer to these entities as chaoticons, as they are termed for the spatiotemporal chaotic localized states in the liquid crystal light valve with feedback loop~\cite{Verschueren-prl-13}. We believe it is the first time, to the best of our knowledge, to present the solutions in the conservative system described by the NNLSE which possess both the chaotic and soliton-like properties.

\emph{Unstable evolution of HGT-SSs}--We consider here the NNLSE [Eq.~(\ref{nnlse})] with the EDRF~\cite{Krolikowski-pre-01, Xu-ol-05, Rasmussen-pre-05}
\begin{equation}\label{response}
R(x)=\frac{1}{2w_{m}}\exp (-\frac{|x|}{w_{m}}),
\end{equation}
which has a singularity at $x=0$. This case corresponds to the model for the propagation of the (1+1)-dimensional paraxial optical beam in nematic liquid crystals~\cite{book-Assanto, guo-book-2015, Conti-prl-03, Conti-prl-04},
in which $q$ is the dimensionless slowly-varying complex amplitude of the optical field, $x$ and ``time'' $t$ stand for, respectively, the dimensionless transverse coordinate and the dimensionless propagation direction coordinate. The relative scale of the characteristic length of the response function $w_{m}$ to the statistic width of the wave $w$ denotes the degree of nonlocality~\cite{Krolikowski-pre-01, Bang-pre-02}, where $w$ is defined by the second-order moment $w(t)=\{2\int_{-\infty}^{\infty}[x-x_c(t)]^{2}|q(x,t)|^{2}dx/P\}^{1/2}$ ($x_{c}(t)=\int_{-\infty}^{\infty}x|q(x,t)|^{2}dx/P$ is the center of the wave and $P=\int_{-\infty}^{\infty}|q(x,t)|^{2}dx$ is the power that is conserved). The larger the ratio $w_{m}/w$, the stronger the nonlocality.

The NNLSE [Eq.~(\ref{nnlse})] permits the SSs of the form~\cite{Buccoliero-prl-07, Xu-ol-05, Dong-pra-10}
\begin{equation}\label{ss}
q(x,t)=u_{N}(x)\exp(ib_{N}t),
\end{equation}
where $u_{N}$ is a real function and $b_{N}$ is a real constant.
It was numerically found that in the case with the EDRF $u_{N}(x)$ is of $N$-humps ($N=1,2,3,...$) HGT-structure~\cite{Xu-ol-05}, specially, $u_{1}(x)$ has a single-hump Gauss-type shape.
It was also proved that $u_{N}(x)~(N\geq2)$ can exist only when the parameters $w_m$ and $b_{N}$ satisfy $w_{m}>1/\sqrt{2b_{N}}$ ~\cite{Rasmussen-pre-05, Yew-jns-99}.

\begin{figure}[htbp]
  \centering
 \includegraphics[width=8cm]{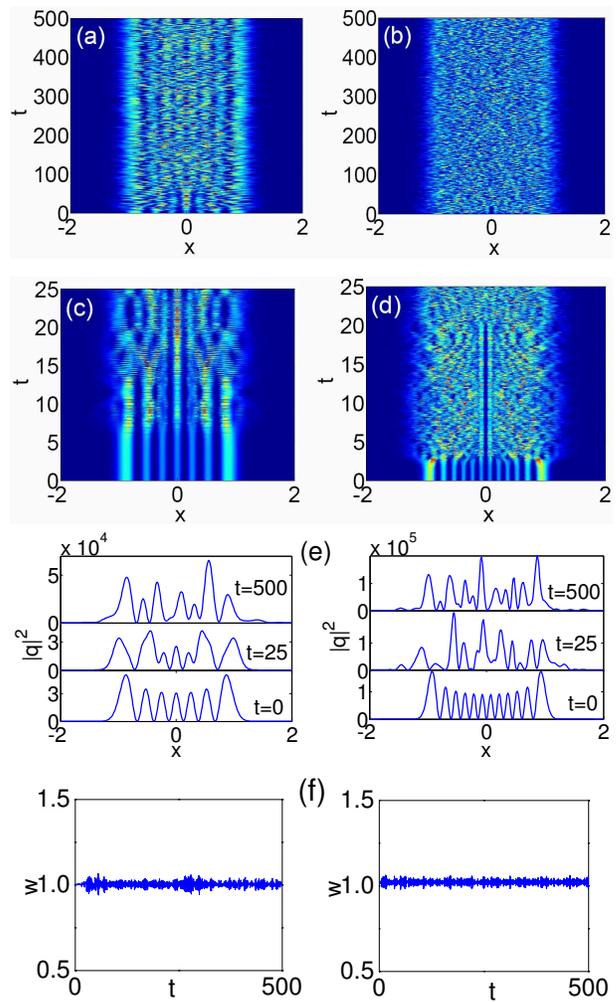}
  \caption{The unstable evolutions of the NNLSE for the initial inputs of the HGT-SSs. (a) and (b): the contour plots for the intensity $|q(x, t)|^{2}$, (c) and (d): the enlargement of the initial region of [0, 25] in (a) and (b), (e): profiles of the intensity at different $t$, (f): the statistic width $w$. The left and right columns are for the SSs with $N=7$ and 12, respectively.}\label{fig-evolution}
\end{figure}

We simulate Eq.~(\ref{nnlse}) and (\ref{response}) with the initial inputs of the HGT-SSs $q(x,0)=u_{N}(x)$ by means of the split-step method~\cite{Agrawal-book-07-2}. The case of strongly nonlocal nonlinearity ($w_{m}=10$ and $w(0)=1$ unless otherwise stated) is considered. The unstable evolution of the HGT-SSs ($N>4$) are given in Fig.~\ref{fig-evolution}, where only solutions with $N=7$ and 12 are displayed without loss of generality.
 It is clear that the profiles starting from regular multi-humps turn to be irregular shapes, several of which are shown in Fig.~\ref{fig-evolution} (e).
The evolution diagrams remind us the behavior of chaos.

\emph{Chaotic behavior: positive Lyapunov exponents}--Since a positive Lyapunov exponent is a signature of chaos, we explore the maximal Lyapunov exponent (MLE) ~\cite{book-Sprott, book-Schuster, Lissauer-rmp-99, Verschueren-prl-13} for the evolution of the HGT-SS. 
According to Refs.~\cite{Cassidy-prl-09, Brezinova-pra-11,Tancredi-aj-01}, the MLE is computed by
\begin{equation} \label{mle}
\lambda=\lim_{r\rightarrow 0}\lim_{t\rightarrow\infty}\frac{1}{t}\ln\frac{d(q_{1}, q_{2}; t)}{d(q_{1}, q_{2}; 0)},
\end{equation}
where $d(q_{1}, q_{2}; t)=[\int_{-\infty}^{\infty}|q_{1}(x, t)-q_{2}(x, t)|^{2}dx]^{1/2}$, which is the distance between two functions $q_{1}(x, t)$ and $q_{2}(x, t)$ 
in the Hilbert space (the $L^{2}$ norm in the Hilbert space), the two initial values $q_{1}(x, 0)=u_{N}(x)$ and $q_{2}(x, 0)=u_{N}(x)+r(x)$, and $r(x)$ is a random perturbation function (as small as machine precision allows).

\begin{figure}[htbp]
  \centering
  \includegraphics[width=5.5cm]{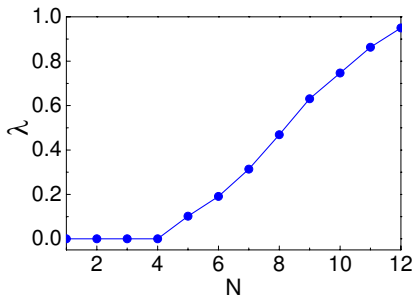}
  \caption{The MLEs for the evolution of the HGT-SSs with different $N$.}\label{fig-mle}
\end{figure}

We have verified that the MLE for every initial input depends neither on $r(x)$ nor on the computing parameters, such as the step size etc.
The MLEs for the HGT-SSs with $N\leq12$ are summarized in Fig.~\ref{fig-mle}. We can see obviously that the MLEs for the unstable SSs are all positive and increase monotonously with $N$, while those for solitons are equal to zero. The occurrence of chaos can be understood as a consequence of the complex interactions among humps. The more humps the profile possesses, the more complex the interactions are, and the higher degree the chaos is of.

It is especially important to make clear that the chaotic phenomenon described above is due to the intrinsic nature of the system but not numerically induced chaos~\cite{Herbst-prl-89}. Although the numerical method applied is not symplectic, it has been demonstrated that this is not the relevant issue for an infinite dimensional Hamiltonian system~\cite{Brezinova-pra-11, Ablowitza-jcp-97}. We have confirmed numerically that the scaling property of the MLEs match the transformation invariance of Eq.~(\ref{nnlse})~\cite{Ouyang-pre-06}. The MLEs for the HGT-SSs with a given number of humps under the condition of $\tilde{w}=w/k$ ($k$ is a positive constant) and $\tilde{w}_{m}=w_{m}/k$ satisfy $\tilde{\lambda}=k^{2}\lambda$ within the error range allowed. The satisfaction of the scaling property is a stringent test for the reliability of numerical computations~\cite{Brezinova-pra-11}.

\begin{figure}[htbp]
  \centering
  \includegraphics[width=5.7cm]{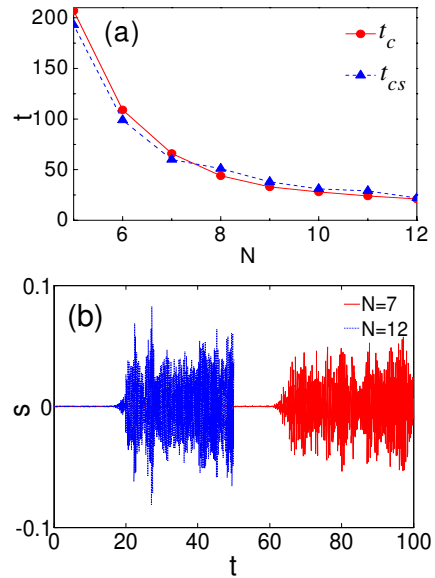}
  \caption{The critical times of becoming completely irregular (a) and the evolution of the skewness (b) for the unstable HGT-SSs.}\label{fig-tc}
\end{figure}

Next, we will prove that the MLEs coincide with the growth rates of the initial numerical errors. Although, literally, the MLE measures the typical exponential rate of growth of an infinitesimal perturbation, the growth of a noninfinitesimal deviation is usually well described in this way. The numerical error of the HGT-SSs computed by the Newton iteration method in double precision is assumed to be of the order of $10^{-9}$. It will make sense that the regular profiles of the initial HGT-SSs will be considered to become completely irregular once the deviation reaches the order of 1.
We can, therefore, estimate $t_{c}$ (the critical time of becoming completely irregular) by $10^{-9}e^{\lambda t_{c}}=1$, thus obtain $t_{c}\approx 20.7/\lambda$, as shown in Fig.~\ref{fig-tc} (a).
From the other aspect, the process of turning to be irregular for the profiles can also be revealed directly in the evolution. Let's consider the skewness (or the third-order central-moment) of the intensity
 \begin{equation}
 s(t)=\frac{1}{P}\int_{-\infty}^{\infty}[x-x_c(t)]^{3}|q(x,t)|^{2}dx.
 \end{equation}
Obviously, there is $s(0)=0$, since $|u_{N}(x)|^{2}$ is symmetric. Fig.~\ref{fig-tc} (b) shows the evolution of $s$ for the HGT-SSs with $N=7$ and 12 in the time interval $[0,100]$ and $[0,50]$, respectively. It can be seen that $|s|$ starts from the close neighbour of zero and then rises abruptly around a certain $t_{cs}$, which is defined as $\int_{0}^{t_{cs}}|s(t)|dt=0.05$. To a great degree, the boom of the skewness indicates the complete irregularity of the intensity profiles. That is to say, $t_{cs}$ represents the critical time of becoming completely irregular attained from the direct statistic method. The comparison between $t_{c}$ and $t_{cs}$ is given in Fig.~\ref{fig-tc} (a). It is evident that the two curves always stay close to each other, which suggests that the critical times evaluated from the above two approaches agree approximately. Then we are certain that the MLEs obtained indeed indicate the exponential growth rates of perturbation.

\begin{figure}[htbp]
  \centering
  \includegraphics[width=8.2cm]{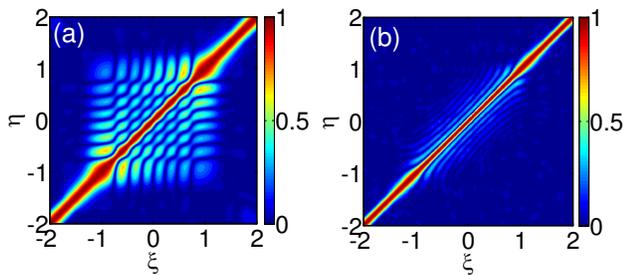}
  \caption{The contour plots of the spatial cross correlation functions for the HGT-SSs with $N=7$ (a) and 12 (b).}\label{fig-xcorr}
\end{figure}

In addition, it is expected that the spatial patterns will be spatially decorrelated in a system described by a partial-differential evolution equation with temporally chaotic behavior~\cite{Cross-science-94, cai-jmp-00, Shlizerman-prl-09, Ramaswamy-sr-16}. Then we calculate the spatial cross correlation function of two long enough wave-amplitude series at locations $\xi$ and $\eta$
\begin{equation} \label{xcorr}
c(\xi, \eta)=\lim_{T\rightarrow\infty}\frac{\int_{t_{0}}^{T}q(\xi,t)q^{*}(\eta,t)dt}{\sqrt{\int_{t_{0}}^{T}|q(\xi,t)|^{2}dt\int_{t_{0}}^{T}|q(\eta,t)|^{2}dt}},~({t_{0}\geq t_{c}}),
\end{equation}
where the superscript $*$ denotes the conjugate complex. The modulus of $c$ for the HGT-SSs are depicted in Fig.~\ref{fig-xcorr}, from which we can see that $|c|$ equals 1 along the line $\xi=\eta$ and decreases rapidly with the separation of two locations. The quick drop of correlation in the $x$ direction means the spatial decoherence~\cite{Shlizerman-prl-09, Ramaswamy-sr-16}.

 \emph{Soliton-like property (I): invariant statistic width}--Chaotic as they are, the evolution of the unstable HGT-SSs maintain almost invariant statistic width $w$, as shown in Fig.~\ref{fig-evolution} (f). The standard deviation of $w$ during the evolution of every HGT-SS is less than 0.02.
 It is well-known that~\cite{Stegeman-science-99, chen-rpp-12, Drazin-1989-book} one of two intrinsic properties for the soliton is its invariant diameter (width), thus we can conclude that the dynamic evolutions of the unstable HGT-SSs with invariant statistic widths are of such a soliton-property from the statistic point of view, even though their profiles during the evolutions are not constant. We can also see, as discussed next immediately, that they still possess the other soliton-property: a particle-like interaction. Because there co-exist the chaotic property and the soliton-like property
 during the dynamic evolution of the unstable HGT-SSs, we refer to them as chaoticons.
 Although we are not first to use the term ``chaoticon'', the intension in both mathematics and physics of the chaoticon here is completely different from that for the spatiotemporal chaotic localized structures in a liquid crystal light valve with optical feedback~\cite{Verschueren-prl-13}.



\begin{figure}[htbp]
  \centering
  \includegraphics[width=7.5cm]{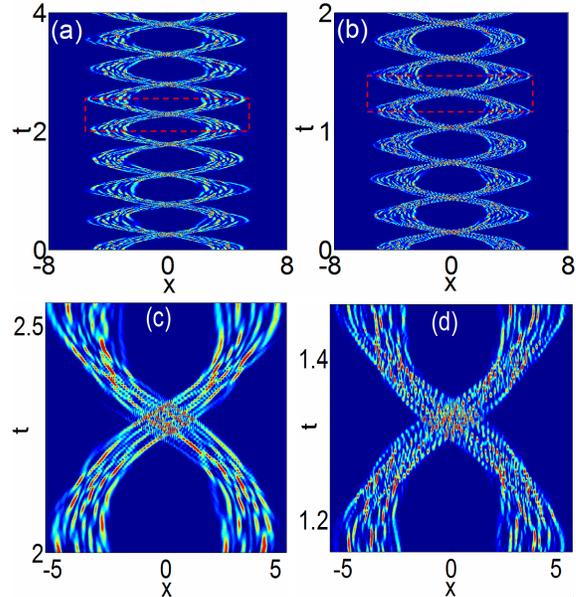}
  \caption{The contour plot of the intensity during the interaction of quasi-elastic collisions between the two chaoticons that are initially identical and paralleled. (a): $N=7$, (b): $N=12$, (c) and (d): partial enlarged details of (a) and (b) in boxes. }\label{fig-interaction}
\end{figure}

\emph{Soliton-like property (II): interaction of quasi-elastic collisions}--Amongst all soliton properties, the important fascinating one is the particle-like interaction~\cite{Stegeman-science-99, Stegeman-ieee-00, Rotschild-np-06}. In order to check whether the chaoticons have such a property, we explore the interaction of two chaoticons that are initially identical and paralleled, which is presented in Fig.~\ref{fig-interaction}. The initial separation between chaoticons is large enough (8 times larger than $w$) to prevent the overlap of waves, and for each case of different $N$s, both of the initial chaoticons are $q(x, t_{0})$ for $t_{0} \geq t_{c}$, which means that the inputs are completely irregular states.
We can observe that the two chaoticons attract each other, and then combine and separate quasi-periodically, much like elastic collisions between two particles. In fact, they will eventually fuse together accompanied by small energy loss to radiation after a much longer evolution. Hence, strictly speaking, the interaction is quasi-elastic.

\emph{Two remarks}--Firstly, it is worth underlining that the evolution of the unstable SSs in generally or weakly nonlocal nonlinearity~\cite{Krolikowski-pre-01} is entirely different from those in strongly nonlocal nonlinearity discussed above. In relatively weak nonlocality, the unstable HGT-SS will break up and form a set of single-hump profiles by emitting remnants of their energy, which is an unbounded state since the radiation waves arrive to infinity~\cite{Kaminer-ol-07}. We have also found that a time when the wave begin to break up increases exponentially with $w_{m}$ for every HGT-SS with a given $w$. It means that the HGT-SSs in stronger nonlocality will evolve longer before they break up. Therefore, the radiation waves are believed to be absent in a strongly nonlocal nonlinear case~\cite{Kaminer-ol-07, Rotschild-np-08}. Secondly, although the system considered is the (1+1)-dimensional NNLSE with the EDRF, our work may be readily extended to systems with different response functions~\cite{Dong-pra-10} or even higher dimensions~\cite{Skupin-pre-06}.

\emph{Conclusions}--We have found that the unstable evolution of the (1+1)-dimensional NNLSE with the EDRF for the initial inputs of the HGT-SSs will evolve into a new kind of chaoticon, which occur only in the case of strongly nonlocal nonlinearity. The chaoticon exhibits both chaotic and soliton-like properties.
The chaotic behavior is signified by the positive MLEs and spatial decoherence. 
The soliton-like property is demonstrated by the invariant statistic width during the evolution, as well as the quasi-elastic collisions during the interaction.

We are grateful to Prof. Senyue Lou at Ningbo University for his great help during the work. This research was supported by the National Natural Science Foundation of China, Grant Nos.~11274125 and 11474109.


\end{document}